\documentclass[11pt,english]{article}
\usepackage{mathptmx}
\usepackage[T1]{fontenc}
\usepackage[latin9]{inputenc}
\usepackage[letterpaper]{geometry}
\usepackage{amstext}
\usepackage{graphicx}
\usepackage{setspace}
\makeatletter
\date{}
\usepackage{cite}

\@ifundefined{showcaptionsetup}{}{%
 \PassOptionsToPackage{caption=false}{subfig}}
\usepackage{subfig}
\makeatother

\usepackage{babel}

\begin{document}

\title{Design and optimization of plasmonic-based metal-dielectric nanocomposite
materials for energy applications}

\author{$^{1,2}$J. Trice, $^{1,2}$C. Favazza, $^{3}$H. Garcia%
\thanks{contact author hgarcia@siue.edu%
}, $^{2,4}$R. Sureshkumar%
\thanks{contact author suresh@wustl.edu%
}, $^{5,6,7}$R. Kalyanaraman%
\thanks{contact author ramki@utk.edu%
}}

\maketitle
\begin{center}
$^{1}$Department of Physics, Washington University, St. Louis, Missouri
63130, USA;
\par\end{center}

\begin{center}
$^{2}$Center for Materials Innovation, Washington University, St.
Louis, Missouri 63130, USA;
\par\end{center}

\begin{center}
$^{3}$Department of Physics, Southern Illinois University, Edwardsville,
Illinois 62026, USA;
\par\end{center}

\begin{center}
$^{\text{4}}$Department of Biomedical and Chemical Engineering, Syracuse,
NY 13203
\par\end{center}

\begin{center}
$^{5}$Materials Science and Engineering, University of Tennessee,
Knoxville 37996, USA
\par\end{center}

\begin{center}
$^{6}$Chemical and Biomolecular Engineering, University of Tennessee,
Knoxville 37996, USA
\par\end{center}
\begin{abstract}
Metallic nanoparticles embedded in dielectrics permit enhanced capture
of light at specific wavelengths through excitation of plasmons, i.e.
the quanta of coherent and collective oscillations of large concentrations
of nearly free electrons. In order to maximize the potential of such
enhanced absorption in useful tasks, such as the generation of carriers
in photocatalysts and semiconductors, it is important to be able to
predict and design plasmonic nanocomposites with desired wavelength-dependent
optical absorption. Recently, a mixing approach formulated by Garcia
and co-workers {[}Phys. Rev. B, 75, 045439 (2007){]} has been successfully
applied to model the experimentally measured broadband optical absorption
for ternary nanocomposites containing alloys or mixtures of two metals
(from Ag, Au or Cu) in SiO$_{\text{2}}$ dielectric. In this work
we present the broadband optical behavior of important optical coating
dielectrics (Si$_{\text{3}}$N$_{\text{4}}$ and SiO$_{\text{2}}$)
and photocatalyst (TiO$_{\text{2}}$) containing various configuration
of nanoparticles of Al, Au, Ag, or Cu. The spectral behavior of various
combinations of the metallic species in the dielectrics was optimized
to show either broadband solar absorption or strong multiple plasmonic
absorption peaks. The applications of such nanocomposite materials
in solar energy harvesting and spectral sensing are also presented
and discussed.
\end{abstract}

\section{Introduction}

The strong optical interaction seen when nanoscale metal-dielectric
structures interact with light promise several applications, including
optical manipulation well below the diffraction limit, high-speed
information processing and integration of Si electronics with optics
\cite{atwater07,quinten98,krenn99,fischer01}. The strong interaction
is a result of the resonant absorption of electromagnetic radiation
by plasmons, which are collective oscillations of large concentrations
of nearly free electrons \cite{Kittel03}. For nanoscale metallic
structures, surface plasmons are primarily responsible for this interaction.
Nanocomposites made from metal nanostructures embedded in or placed
on dielectrics offer significant potential as materials showing enhanced
or tunable light harnessing \cite{StuartAPL98,SchaadtAPL05,PillaiAPL06,ColeAPL06}.
For instance, Schaadt and co-workers \cite{SchaadtAPL05} have shown
that incorporation of Au nanoparticles on the surface of a Si photodiode
results in enhanced photocurrent at specific wavelengths that were
correlated to the surface plasmon resonance. However, the fabrication
of practically useful devices that utilize plasmonics to achieve a
desired optical and/or electronic response requires the careful design
and fabrication of nanomaterials with well defined shape, size and
metal-dielectric composition. Recently, there has been tremendous
activity in controlling the physical attributes of nanoscale metallic
structures, such as the particle size and shape, through a variety
of techniques, including ion irradiation \cite{MishraAPL07,BattaglinNucInsMethPhysB00,MagruderNonCrystSold94,MeldrumNucInsMethPhysRes2001},
laser-assisted chemical vapor decomposition \cite{ElihnJAP07,ElihnAppPhysA01}
and laser-induced dewetting \cite{Trice07a,GangopadhyayNanoTech07,Favazza06b,Favazza05b}.
However, the fabrication of metal-dielectric nanocomposites driven
by a knowledge-based design of optical properties is very important
towards realizing cost-effective means for manufacturing such novel
materials. 

In this work we focus on a theory-driven design and optimization of
metal-dielectric nanocomposites that can show broadband optical absorption
via plasmonic effects. The desire to achieve broadband plasmonic absorption
comes from the various experimental results that show that nanoparticles
and nanocomposite coatings on photovoltaic devices can give rise to
a wavelength-dependent photocurrent enhancement that correlates to
the plasmonic responses of the applied material \cite{SchaadtAPL05,DerkacsAPL06,PillaiAPL06,StuartAPL98}.
While consideration of the effects of electromagnetic scattering due
to the absorbing-layer's microstructure leads to further optimization
\cite{DerkacsAPL06,PillaiJAP07}, optimizing the absorption of incident
solar energy though judicious design of metal type, nanoparticle shape,
its volume fraction, and the host dielectric \cite{ZhangAPL07,DouillardJAP07,SuAPL07,StuartAPL98}
is likely to be the first step to an improved device. Recently, a
homogenization procedure has been proposed using the tight lower bounds
of the Bergman-Milton formulations to characterize the optical properties
of ternary nanocomposites consisting of metallic nanoparticles embedded
in various host dielectrics \cite{GarciaPRB07}. Using this model,
excellent agreement was found between the experimentally measured
and theoretically estimated optical absorption for a variety of plasmonic
systems, including binary alloys, mixtures and core-shell structures.
Based on this success, we have extended the mixing approach and applied
it to the case of quaternary metal-SiO$_{\text{2}}$ nanocomposites,
i.e. a system containing three different metals in SiO$_{\text{2}}$,
to predict the broadband spectral response. We showed that a judicious
selection of metal type, their concentration and particle size can
result in broadband plasmonic light absorption \cite{TriceSPIE07a},
a feature that could be very beneficial in applications that need
to maximize the absorption of incident broadband solar radiation,
such as in solar cells and photodetectors . 

Here we present calculations of the broadband optical properties of
other useful dielectric material, including Si$_{3}$N$_{4}$ and
TiO$_{\text{2}}$. Si$_{\text{3}}$N$_{\text{4}}$ is used in conventional
solar cell technology as an anti-reflective coating layer \cite{SoporiJElecMat2003}.
We have calculated the optical properties of binary and quaternary
nanocomposites (NC) made from different metals (Al, Cu, Ag, Au) in
Si$_{\text{3}}$N$_{\text{4}}$ and determined wavelength regimes
of strong absorption. Based on these results we have designed a broadband
absorbing nanocomposite whose absorption spectra is optimized to match
the shape of the 1.5 AM incident solar radiation. Our results show
that a quaternary NC containing Cu, Ag and Al in Si$_{\text{3}}$N$_{\text{4}}$
shows extremely good broadband absorption spectra. In comparison,
similar broadband absorption in SiO$_{\text{2}}$ requires Cu, Ag
and Au nanoparticles. TiO$_{2}$ is an essential component in dye-sensitized
solar cells \cite{GratzelNat1991} and its photocatalytic properties
have a variety of uses, including in paint, in self-cleaning glasses,
and in multifunctional coatings \cite{BrookSurfCoatTech2007}. However,
one of the outstanding challenges in TiO$_{\text{2}}$ is to enhance
its absorption in the visible spectrum. Towards this goal, we have
explored the plasmonic absorption of several different metals (Al,
Cu, Ag, Au) in TiO$_{\text{2}}$. Our calculations show that strong
absorption occurs in the range of 670-700 nm by Cu, Ag or Au while
Al shows broad absorption from the UV into the visible till about
500 nm. The results of this work show that it is possible to design
and optimize the optical behavior of metal-dielectric nanocomposites
with desired optical response, including for applications in energy
harvesting. The remainder of this paper is divided as follows. In
Sec. \ref{sec:Homogenization-technique} we discuss the basic ideas
behind plasmonics and plasmonic absorption and present the homogenization
scheme that permits a theoretical estimate of the optical absorption
in multimetal-dielectric nanocomposites. In Sec. \ref{sec:Potential-Applications}
we discuss potential applications resulting from NCs that show absorption
in broadband and/or at specific wavelengths. We also discuss the computational
optimization scheme based on the simulated annealing technique that
permits the optimization of a nanocomposite. We conclude the paper
with brief discussions in Sec. \ref{sec:Conclusions}.

\section{Homogenization Scheme \label{sec:Homogenization-technique}}

The essential physics of the plasmonic properties of metals irradiated
by electromagnetic radiation in the optical regime are captured by
the Drude-Sommerfeld theory of metals, with corrections to the kinetics
of the free electron Fermi gas due to energetic contributions from
bound electrons \cite{AshcroftSSPhys1976}. When a collection of atoms
are brought together forming a crystalline lattice, one must determine
the energy states of the \emph{system} of valence electrons as their
motion is influenced by the lattice ions and mutual interaction. In
practice, the wave functions of the valence electrons are not strongly
concentrated about their original atomic position but spread wide
through the metal with a distance comparable to the mean-free path
of the electrons ($\sim50$ angstroms). Hence, the valence electrons
form a sea filling the various electronic states of the system as
a whole \cite{ZimanElecPhon1960}. A plasma oscillation in a metal
is a collection of longitudinal or transverse excitation of these
free electrons. A plasmon refers to the quanta of such an oscillation
which may be stimulated by an incident electromagnetic (EM) wave.
Preferential coupling between an incident EM wave and a particular
plasmon mode can give rise to resonance phenomenon. At normal incidence,
for EM wavelengths corresponding to this resonance frequency, the
momentum of the plasmon will be negligible because the momentum of
the reflected photon is virtually zero. Thus, excitation of the plasmon
mode is localized at the surface of the medium. However, for certain
angles, when requirements for energy and momentum conservation are
met, a surface-plasmon polariton wave may be induced that propagates
along the surface of the media. The existence of these self-sustaining
collective excitations at metal surfaces was originally predicted
by Ritchie \cite{RitchiesPhysRev1957}.For the case of metallic nanoparticles,
surface plasmonic effects are the dominate mechanism dictating the
interaction with EM energy. When metallic nanoparticles are embedded
in a host dielectric, the precise wavelength where EM energy and plasmonic
modes couple at metal-film interface is dictated by the dielectric
coefficient for the individual phases. The essential physics of this
charge oscillation for dilute systems of metal nanoparticles ($<10\%$)
in a dielectric matrix is captured by the lower bound for the effective
dielectric coefficient proposed by Bruggeman \cite{BergmanPRB1981}
and Milton \cite{MiltonJAP81}. We have taken the binary dielectric
mixing rule from Bruggeman and Milton and extended its application
to multi-metal nanocomposite systems via a homogenization technique
\cite{GarciaPRB07}. 

\begin{figure}[t]
\subfloat[]{

\includegraphics[width=3in]{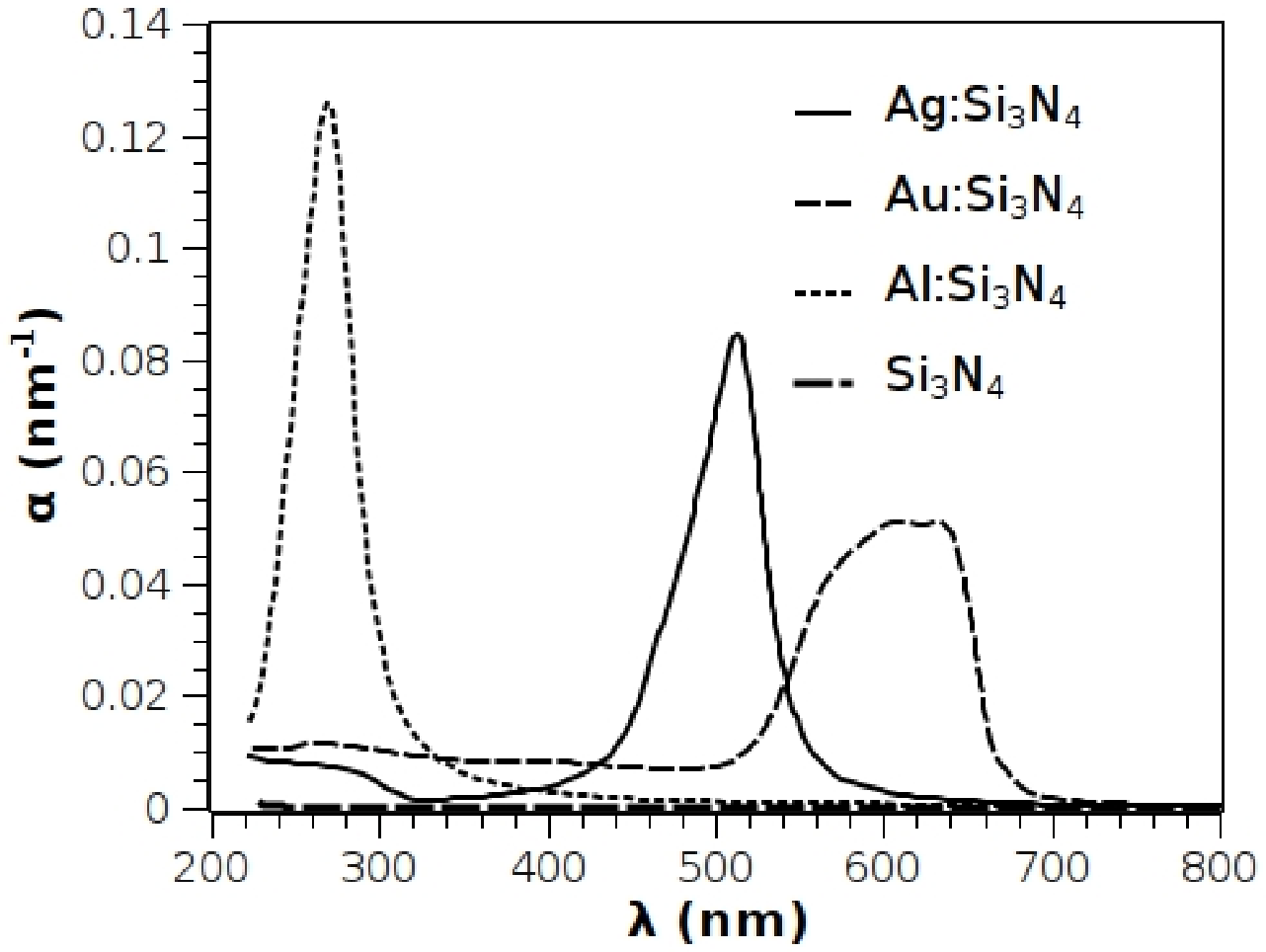}}\subfloat[]{

\includegraphics[width=3in]{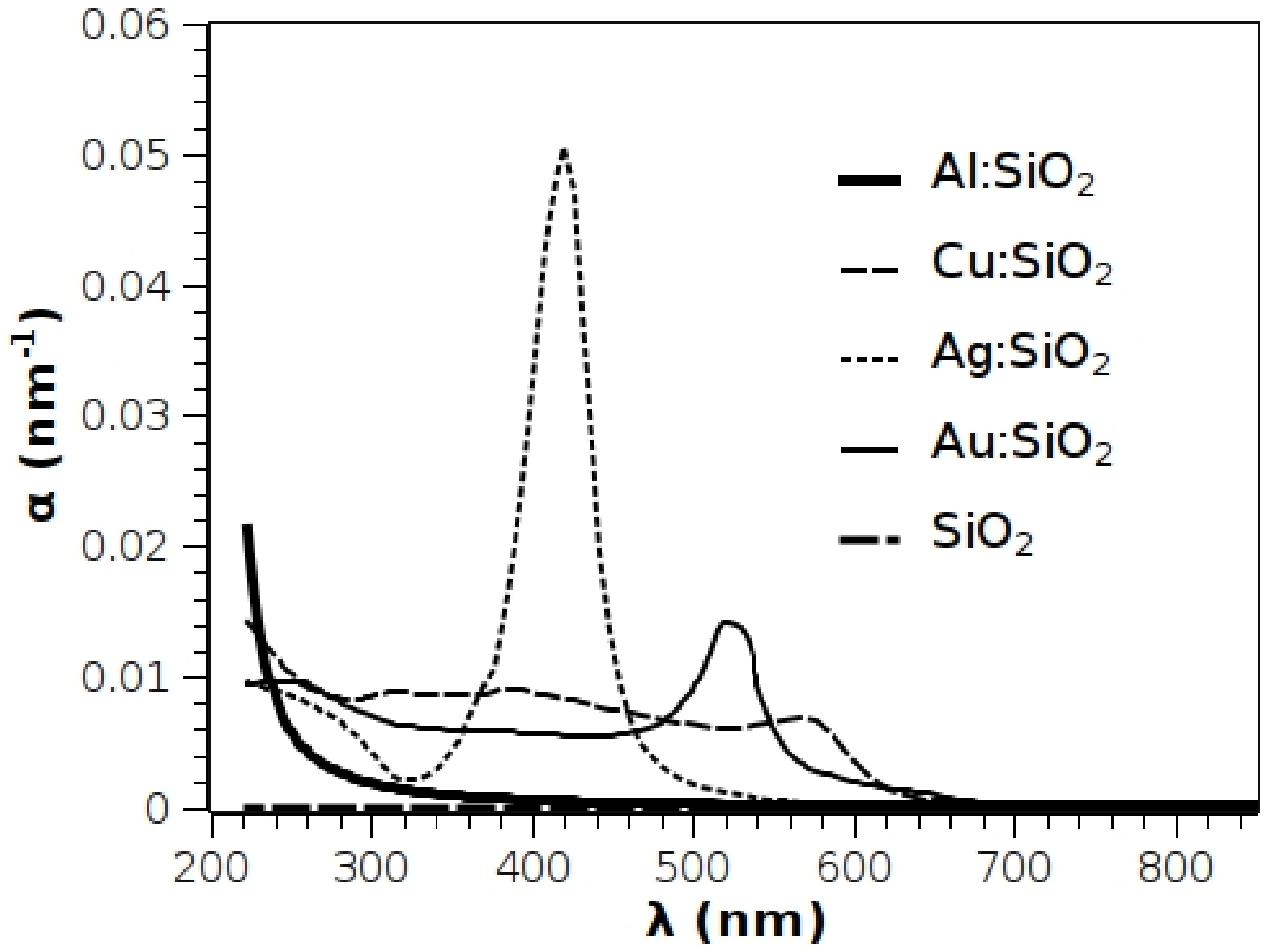}}

\subfloat[]{

\includegraphics[width=3in]{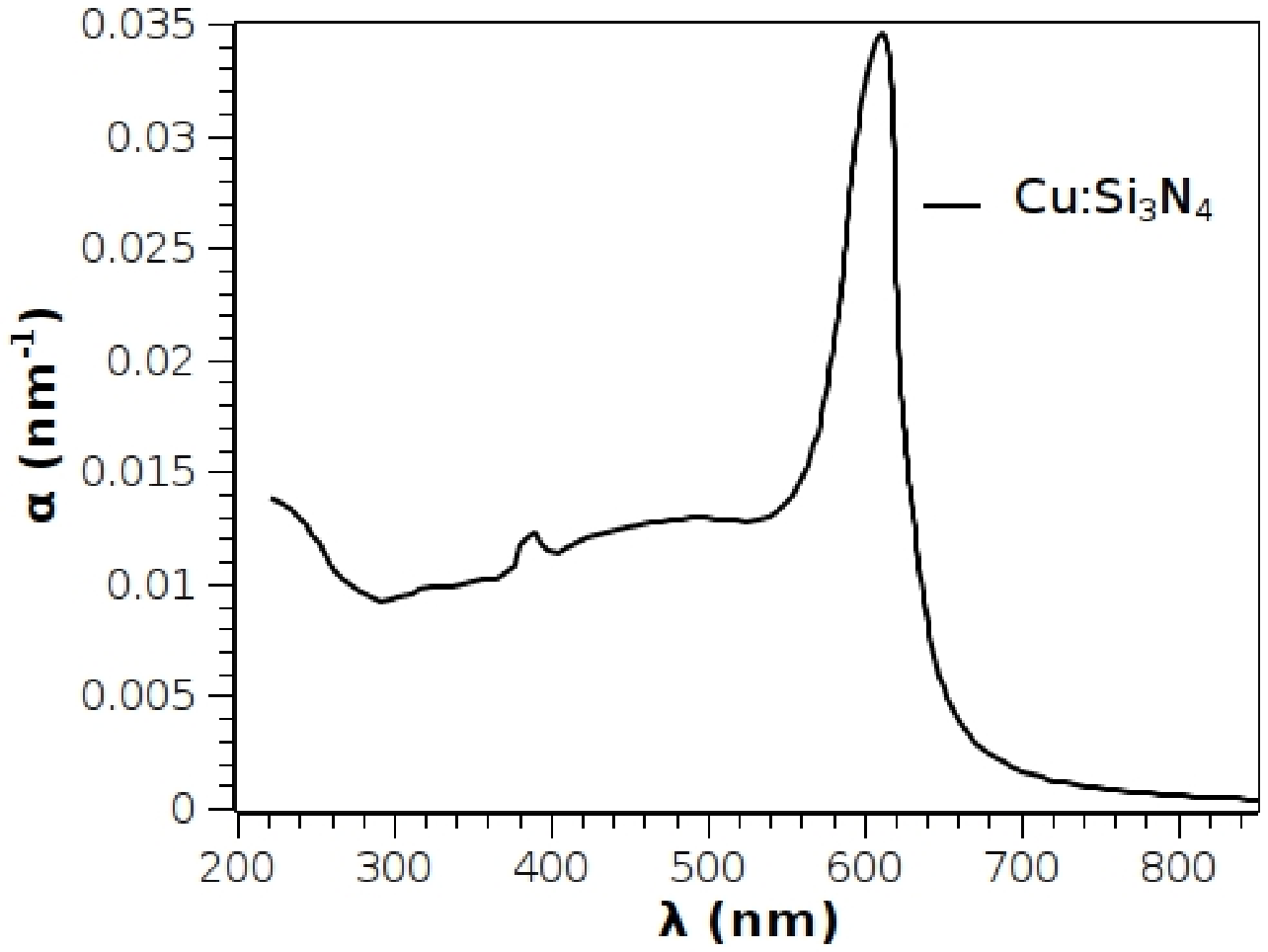}}\subfloat[]{

\includegraphics[width=3in]{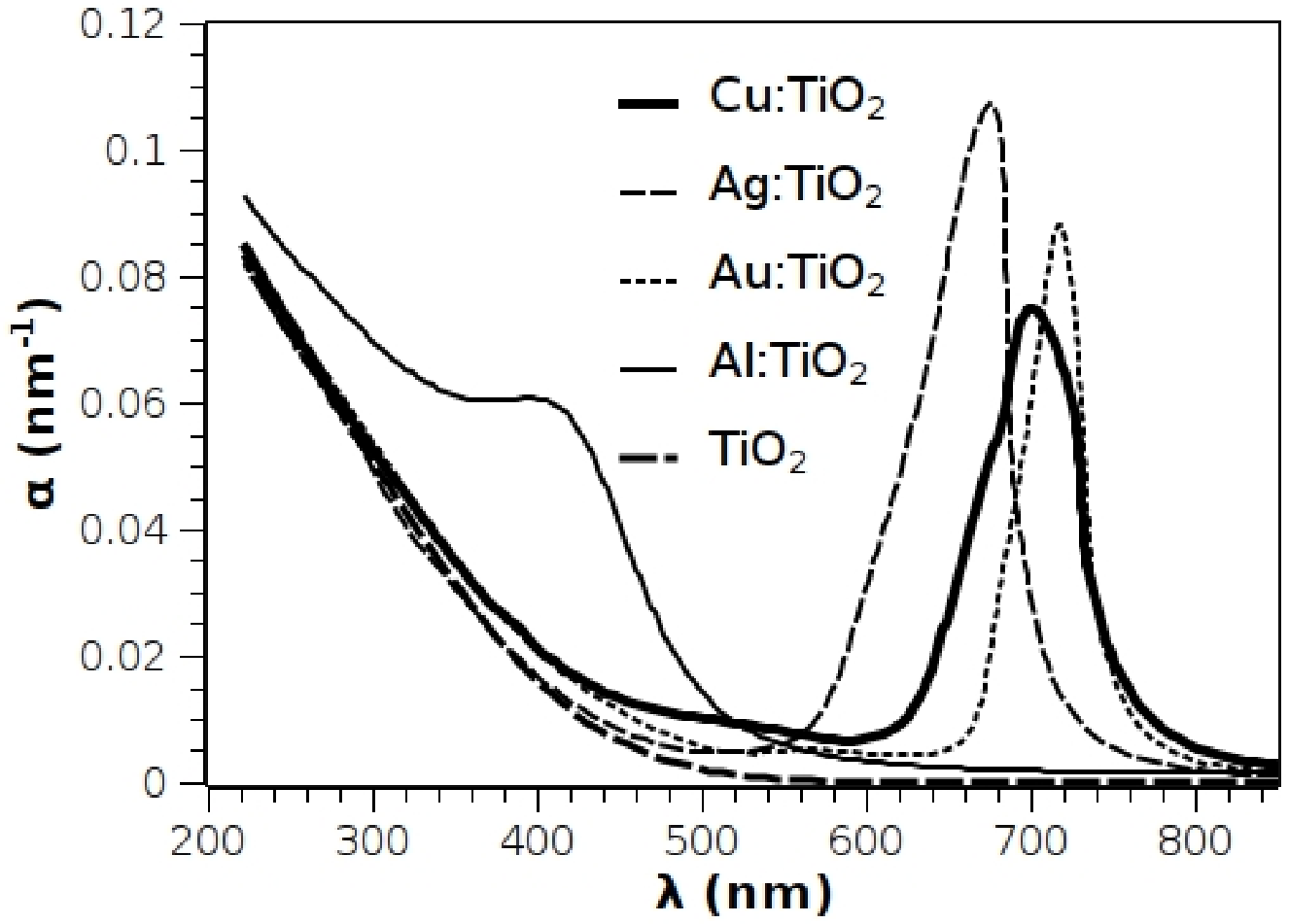}}

\caption{\label{fig:Singlemetals}The absorption coefficient of single species
metal inclusions embedded in host dielectrics over the spectral range
of 200 - 850 nm. Here, size effects of the inclusions are not considered.
(a) displays the absorption of Al, Ag, and Au spheres in Si$_{3}$N$_{4}$.
The strongest absorption enhancement is localized near a particular
wavelength (namely $268$ nm, $513$ nm, and $633$ nm for Al, Ag,
and Au respectively). (b) shows the absorption of Al, Ag, and Au in
SiO$_{2}$. Ag and Au exhibit strong plasmonic responses near 414
nm and 529 nm respectively. Al in SiO$_{2}$ shows strong absorption
in the UV near the edge of the available experimental data. (c) presents
the absorption spectra of Cu spheres embedded in Si$_{3}$N$_{4}$.
The system exhibit a broad response spanning from the UV well towards
the visible regime. A very strong absorption enhancement is noted
near $\sim640$ nm for the case of Cu embedded in Si$_{3}$N$_{4}$.
(d) demonstrates the absorption coefficient spectrum of Al, Ag, Au
and Cu in TiO$_{2}$. Cu, Au and Ag all exhibit strong enhancements
localized near $701$ nm, $712$ nm, and $672$ nm respectively. Al
in TiO$_{2}$ exhibits a broad response ranging from the UV to a wavelength
$\sim500$ nm. (a), (b) and (d) also contain plots of the absorption
spectrum of the host dielectric without metal inclusions for comparison.
}

\end{figure}

In the limit of long electromagnetic wavelength relative to the system
microstructure, the technique of homogenization, where two or more
constituent phases are blended together in such a manner that the
resulting composite material appears as a continuous material, is
an important tool in characterising the desired macroscopic responses
of engineered materials. A large amount of research is dedicated towards
placing bounds on the maximum and minimum values of the macroscopic
dielectric coefficient that composite materials may take when only
limited information is provided on features of the constituent phases
including: volume fraction, average spacing, and/or spatial distribution
\cite{ShenJAP1990,BermanPRL80,BergmanPRB1981,MiltonJAP81,Sihvola99}.
Such models have been used in geological applications and are useful
in predicting the behavior of natural occurring composites such as
reservoir rocks \cite{ShenJAP1990} or ice \cite{Sihvola99}, which
may contain dilute concentrations of useful resources such as natural
gas, water, or minerals. While many bounds are available in the literature,
certain bounds are more useful towards narrowing down the range of
predicted macroscopic behavior for a particular situation of interest
\cite{Sihvola99}. For example, the Wiener bounds are the loosest,
giving the maximum and minimum values of the effective dielectric
function independent of the type of mixture in question. For systems
that are isotropic and with uniformly distributed constituents, the
Hashin-Shtrikman bounds, based on a variational formulation of the
energy functional for the mixture, represents a tighter range of values
than the Weiner bounds at the expense of the added assumptions \cite{KarkkainenIEEE2000}.
In addition to the Hashin-Shtrikman approach, two other widely circulated
bounds in the literature are the Maxwell-Garnett and Bruggeman formalisms
\cite{DuncanOptComm2007}. However, these tighter approaches do not
necessarily yield physically realistic estimates in certain parameter
regimes \cite{MackayOptComm2004}. Recently we have shown that a homogenization
procedure based on the tight lower bound given by Bergman and Milton
\cite{BermanPRL80,MiltonJAP81} is able to predict optical data in
dilute nanocomposite materials without the use of any fitting parameters
\cite{GarciaPRB07}. In this procedure, the effective dielectric function
for a binary metal-dielectric composite can be expressed as follows:

\begin{equation}
\epsilon_{eff}(\gamma)=\left[\frac{f_{a}}{\epsilon_{a}}+\frac{f_{h}}{\epsilon_{h}}-\frac{2f_{a}f_{h}\left(\epsilon_{a}-\epsilon_{h}\right)^{2}}{3\epsilon_{a}\epsilon_{h}\left[\epsilon_{h}\gamma+\epsilon_{a}\left(1-\gamma\right)\right]}\right]^{-1}\label{eq:BergMilt}\end{equation}
where $f$ denotes the component's volume fraction, $\epsilon$ is
the component's dielectric function depending on the wavelength of
the incident electromagnetic energy with subscripts $a$ and $h$
referencing the metal and host matrix respectively. $\gamma$ is a
geometrical factor taking into account the shape of the metal particles
which may posses spatial dependence for non-symmetric inclusion geometries
\cite{Sihvola99}. For the purposes of this paper we are interested
in spherical particles and accordingly choose $\gamma=\frac{2}{3}(1-f_{a})$.
Here, we also assume the spherical inclusions are distributed with
enough uniformity in the host matrix that system spatial dependence
may also be neglected. Equation \ref{eq:BergMilt} represents the
effective dielectric function for the two component case of a metal
particle embedded in a host dielectric matrix. Note it is customary
to let $f_{h}=1-\sum f_{n}$ where $f_{n}$ corresponds to the metal
components \cite{Sihvola99}. Our analysis made use of experimentally
measured optical data from Ref. \cite{SopraDataBase}. In Fig. \ref{fig:Singlemetals}
the effective absorption coefficient $\alpha$, given by \begin{equation}
\alpha=\frac{2\sqrt{2}\pi}{\lambda}(\sqrt{Re(\epsilon_{eff})^{2}+Im(\epsilon_{eff})^{2}}-Re(\epsilon_{eff}))^{1/2}\label{eq:absorptionComplete}\end{equation}
 where $\lambda$ is the wavelength of incident electromagnetic energy,
of a single metal species ($f_{1}=10\%$) embedded in a Si$_{3}$N$_{4}$,
TiO$_{2}$, or SiO$_{2}$ matrix is presented. For the case of Al,
Au, and Ag nanoparticles in Si$_{3}$N$_{4}$, the composite system
possesses a vary sharp increase in absorption localize about a particular
wavelength ( $268$ nm, $513$ nm, and $633$ nm for Al, Ag, and Au
respectively) (Fig. \ref{fig:Singlemetals}a). Similar behavior is
observed for Au, Ag, and Cu (Fig. \ref{fig:Singlemetals}d) nanoparticles
embedded in TiO$_{2}$ with localized absorption enhancements $712$
nm, $672$ nm, and $701$ nm respectively, all corresponding to photons
with energies below the bandgap of TiO$_{2}$ (3.0 eV). Similarly,
Ag and Au in SiO$_{2}$ also exhibit strong absorption near their
plasmonic responses with respective absorption enhancements in the
vicinity of 414 nm and 529 nm, respectively (Fig. \ref{fig:Singlemetals}b).
Al in SiO$_{2}$ shows strong absorption in the UV near the edge of
the available experimental data (Fig. \ref{fig:Singlemetals}b). However,
Cu inclusions in Si$_{3}$N$_{4}$ and SiO$_{2}$ exhibit a broadband
response spanning from the UV to the visible regime (Figs. \ref{fig:Singlemetals}b
and \ref{fig:Singlemetals}c). Near 640 nm Cu in Si$_{3}$N$_{4}$
possesses a very strong absorption enhancement. It is of interest
to note that the resonance response of Cu, Au, and Ag in Si$_{3}$N$_{4}$
all correspond to photon energies below the bandgap of Si$_{3}$N$_{4}$
(4.5 eV). In addition, Al in TiO$_{2}$ exhibits a broader response
than the other presented metals embedded TiO$_{2}$ with an absorption
enhancement ranging from the UV to a wavelength $\sim500$ nm (Fig.
\ref{fig:Singlemetals}d). In Fig. \ref{fig:Singlemetals}(a), \ref{fig:Singlemetals}(b),
and \ref{fig:Singlemetals}(d) the absorption spectrum of the dielectric
material with no metal inclusions is also shown for comparison purposes.
Note that the presence of metal nanoparticles greatly enhances the
absorption predicted from that of the standalone dielectric matrix. 

We now discuss the optical behavior of dielectric containing nanoparticles
of multiple different metals. For the systems considered here, the
analysis is limited to the dilute case (i.e. total volume fraction
of metal versus dielectric $\leq10\%$) because it is desirable from
a practical standpoint of expense and fabrication efficiency. An additional
constraint we have placed is to set the minimum nanosphere diameter
to be no less than $10$ nm. Above this size regime, exotic quantum
confinement effects may be neglected \cite{HalperinRevModPhys1986}.
Furthermore, nanoparticles of size $>$ 10 nm can be assembled on
many dielectric substrates using robust manufacturing techniques such
as those mentioned in the introduction. Figures illustrating the hierarchical
mixing process for both 2- and 3-metal mixtures have been presented
elsewhere \cite{GarciaPRB07,TriceSPIE07a}. The top level represents
the effective dielectric function for the target nanocomposite of
design interest. At each level, the total volume fraction is constrained
such that $f_{h}+\sum f_{n}=1$. During the mixing process, the average
electric field within the composite is held constant while the final
effective permittivity is calculated using equal volumes at each level
of mixing. As described in Ref. \cite{GarciaPRB07}: the effective
permittivity of an N-component mixture can be determined by mixing
N-1 binary mixtures, each comprising of a host and a distinct metal,
the host being common to the N-1 pairs.

At the plasmonic resonance, the absorption is not infinite in extent
and broadened absorption lines are observed due to the finite relaxation
times of scattering \cite{ZimanPrincSolids1972}. For the case of
spherical metal particles of nanometer extent, finite size corrections
are introduce through modification of the imaginary portion of the
effective dielectric coefficient. Eq. \ref{eq:teff} quantifies the
effective change in the electron relaxation time due to scattering
of electrons on the metal-dielectric interface as a slight modification
to the bulk electron relaxation time \cite{YangApplPhysA96} expressed
by : 

\begin{equation}
\frac{1}{\tau_{eff}}=\frac{1}{\tau_{bulk}}+\frac{\nu_{F}}{2d_{n}}\label{eq:teff}\end{equation}
where $\tau_{eff}$ is the effective electron relaxation time, $\tau_{bulk}$
is the bulk electron relaxation time, $d_{n}$ corresponds to the
particle's diameter and $\nu_{F}$ is the velocity of the electrons
near the Fermi surface. The values of $\tau_{bulk}$ and $\nu_{F}$
for the various metals examined in this study were taken from Ref.
\cite{AshcroftSSPhys1976}. The size effect may be incorporated in
the mixing approach though modification of the imaginary portion of
the effective dielectric coefficient \cite{GarciaPRB07} as follows

\begin{equation}
Im(\epsilon_{a})=\frac{\omega_{p}^{2}}{\omega^{3}\tau_{eff}}=Im(\epsilon_{a}^{bulk})\left(\frac{2d_{n}+\nu_{F}\tau_{bulk}}{2d_{n}}\right)\label{eq:ImagCorr}\end{equation}
where $\omega_{p}$ is the plasmon frequency and $\omega$ is the
angular frequency of the incident electromagnetic energy. By manipulating
this nanosize effect, one is able to broaden the plasmonic peaks at
the expense of their magnitude \cite{GarciaPRB07,TriceSPIE07a}.

\section{Potential Applications and Optimization Scheme \label{sec:Potential-Applications}}

\begin{figure}[th]
\subfloat[]{

\includegraphics[height=2.5in]{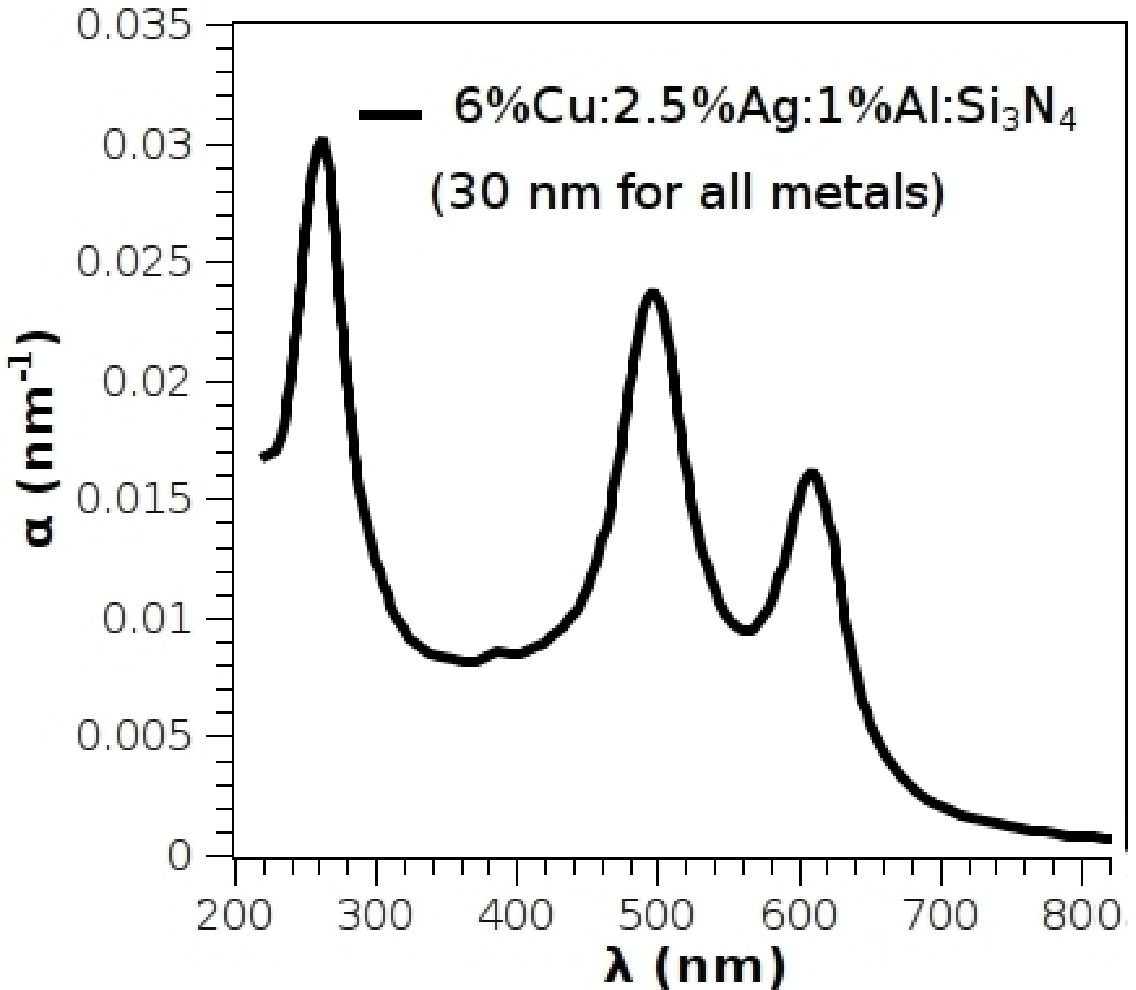}}\subfloat[]{

\includegraphics[height=2.5in]{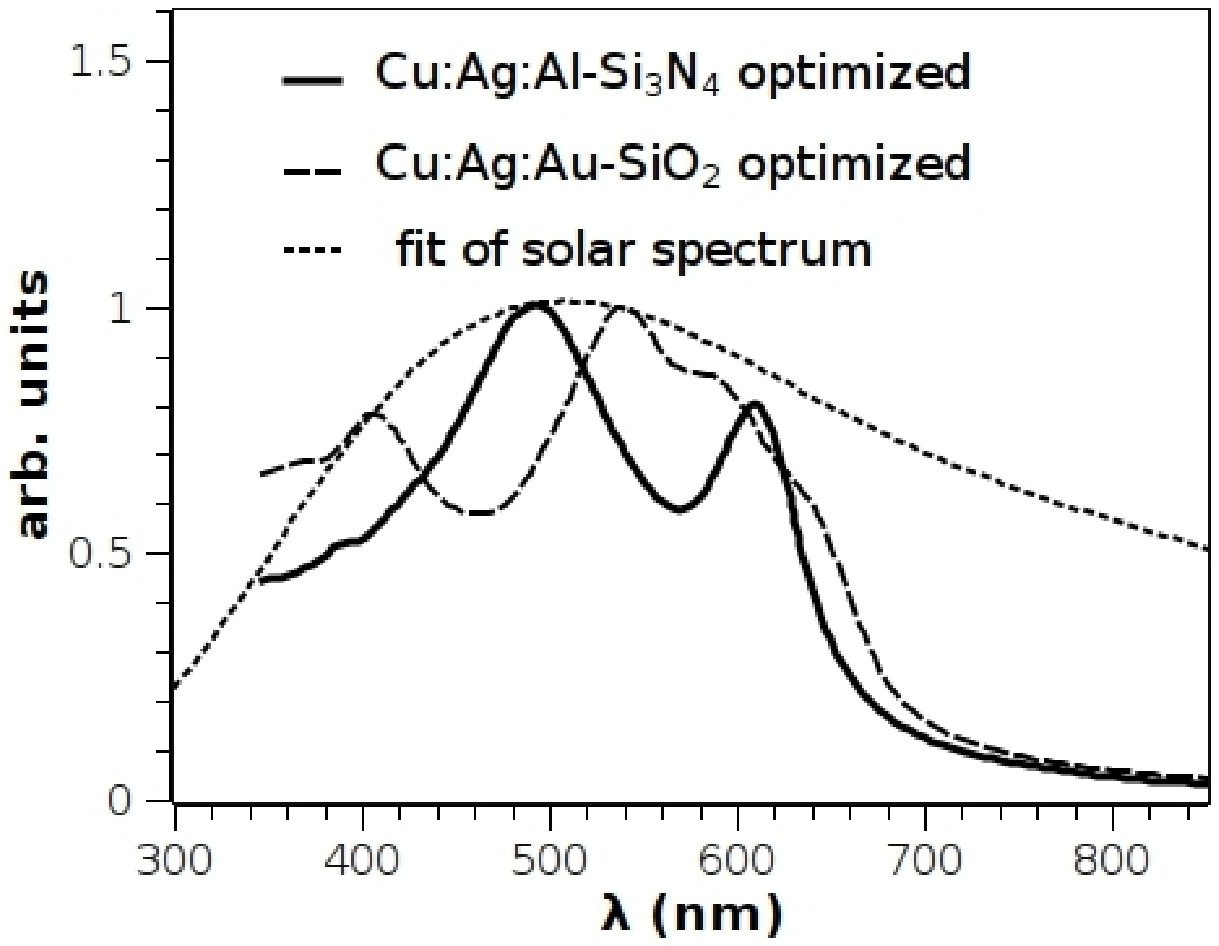}}

\caption{\label{figAppl} Potential designs for applications in spectral sensing
(a) and solar energy harvesting (b). (a) presents a system configuration
where the spectral response is broadband with absorption peaks at
264 nm, 496 nm, and 609 nm due to the respective interaction of Al,
Ag, and Cu with Si$_{3}$N$_{4}$. The chosen configuration (6.0\%
Cu, 2.5\% Ag, and 1.5\% Al with all spherical inclusions possessing
a diameter of 30 nm) causes the maximums of the peaks to descend over
the spectral range. (b) presents the solution for a Cu:Ag:Al-Si$_{3}$N$_{4}$
nanocomposite with absorption coefficient optimized to the solar spectrum
using a simulated annealing algorithm. The configuration was determined
to be 5.45\% Cu, 0.421\% Al, and 4.11\% Ag in Si$_{3}$N$_{4}$ with
respective nanosphere diameters of 30 nm, 22.27 nm, and 10.0 nm. The
optimized as presented and with numerical parameters as discussed
in Ref. \cite{TriceSPIE07a} for the case of a Cu:Ag:Au-SiO$_{2}$
nanocomposite is shown as well. The optimal configuration determined
in that study was 1.80\% Cu, 0.35\% Ag, and 6.40\% Au with respective
particle diameters of 10.1 nm, 29.6 nm and 10.0 nm.}

\end{figure}

\begin{figure}[t]
\subfloat[]{

\includegraphics[height=2.5in]{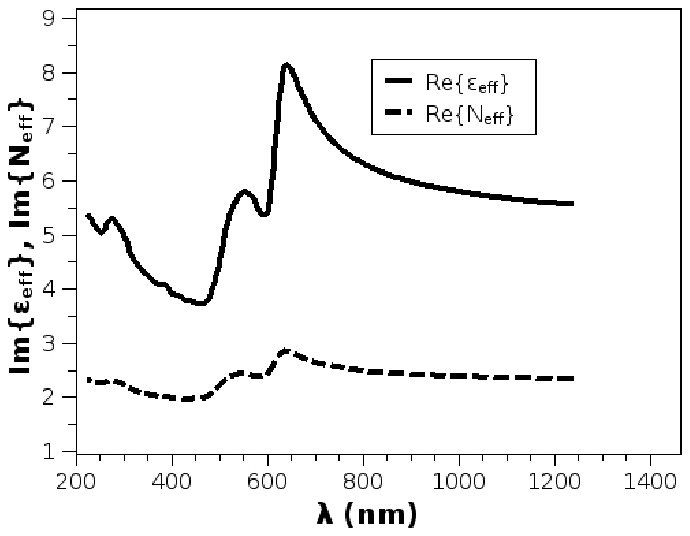}} \subfloat[]{

\includegraphics[height=2.5in]{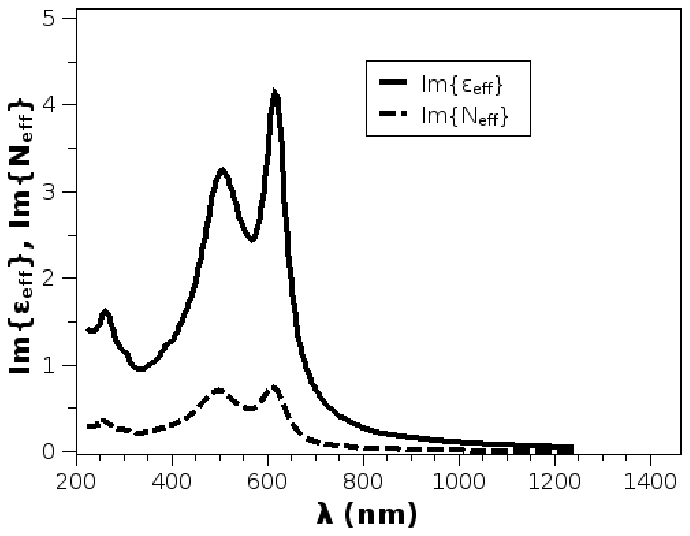}}

\caption{\label{fig:CxEeffNeff}Complex dielectric constant $\epsilon_{eff}$
and index of refraction $N_{eff}$ for a 5.85\% Cu, 0.42\% Al, and
3.71\% Ag in Si$_{3}$N$_{4}$ nanocomposite. }

\end{figure}

In Fig. \ref{figAppl}, the optical absorption response obtained by
the mixing approach are presented with specific applications in consideration.
We begin by noting that it has been previously shown that an absorbing
layer on a photovoltaic device can exhibit photocurrent enhancement
correlating with the plasmonic responses of the layer \cite{SchaadtAPL05,BostromSolarEngMat2004}.
This motivates the application discussion here. Fig. \ref{figAppl}(a)
presents a system configuration where the spectral response is broadband
with absorption peaks at 264 nm, 496 nm, and 609 nm due to the respective
interaction of Al, Ag, and Cu with Si$_{3}$N$_{4}$. The chosen configuration
(6.0\% Cu, 2.5\% Ag, and 1.5\% Al with all spherical inclusions possessing
a diameter of 30 nm) causes the maximum of the peaks to occur at different
locations over the spectral range. While optimizing the size and concentration
can result in several different absorption profiles \cite{TriceSPIE07a},
Fig. \ref{figAppl}(a) illustrates an important design example, i.e.
one in which absorption can be simultaneously maximized at several
different and specific wavelengths. In Fig. \ref{figAppl}(b) we present
the design solution for a Cu:Ag:Al-Si$_{3}$N$_{4}$ nanocomposite
whose absorption coefficient is optimized to resemble the shape of
the incident solar spectrum (1.5 AM) using a simulated annealing algorithm
\cite{MetropolisJChemPhys1953,KirkpatrickScience1983,PressNumRecC1997}.
We define the {}``energy'' to be minimized as $E(\mathbf{f},\mathbf{d})=\sum abs\left|(s_{i}-\alpha_{i}(\mathbf{f},\mathbf{d}))\right|$
with composite parameters $\mathbf{f=}f_{1},f_{2}..f_{n}$ and $\mathbf{d=}d_{1},d_{2}..d_{n}$.
$s_{i}$ and $\alpha_{i}$ represent the normalized solar spectrum
data \cite{RenewableDataCen} and effective nanocomposite absorption
at a given wavelength $i$ respectively. We define this {}``energy''
for $i$ spanning from 200 to 900 nm. This choice of $E$ allows for
the absorption profile that most closely resembles the shape of the
solar spectrum to be determined. The algorithm begins by initializing
the system to some initial state $E^{s}(\mathbf{f},\mathbf{d})$.
A neighboring state $E^{n}$ is called by using the condition\begin{equation}
E^{n}(\mathbf{f}^{n},\mathbf{d}^{n})=E^{s}(\mathbf{f}+\Delta\mathbf{f},\mathbf{d}+\Delta\mathbf{d})\label{eq:neighbor}\end{equation}

where $\Delta\mathbf{\mathbf{f}}=f_{max}\mathbf{x}$ with $\mathbf{x}$
being a vector of dimension $n=3$ with each component taking a random
value between $\left\{ -1,1\right\} $. $f_{max}$ represents the
maximum magnitude the volume fraction of a particular species may
step. Similarly, $\Delta\mathbf{\mathbf{d}}=d_{max}\mathbf{x}$ with
$d_{max}$ representing the maximum magnitude the diameter of the
particles corresponding to a particular species may step. Here, values
of $f_{max}=0.1\%$ and $d_{max}=0.1$ nm provided sufficient finesse
in moving through the search space. Next, the neighboring state is
compared with the best state encountered thus far $E^{b}$ (where
$E^{b}$ was originally initialized to the same state as $E^{s}$).
If $E^{n}<E^{b}$ then $E^{b}$ is set equal to $E^{n}$. Then, the
algorithm must decide if this neighbor state will become the preferred
state for the system. To accomplish this, a Boltzmann-type probability
$P$ analogous to classical statistical physics is calculated-- namely
\begin{equation}
P=\exp(\frac{-(E^{n}-E^{s})}{kT})\label{eq:prob}\end{equation}

where $T$ is the annealing parameter analogous to temperature and
$k$ is a constant used to refine the annealing schedule. Note that
if $P>1$ then $P$ is simply reassigned to 1. $P$ is then compared
to a random number $x$ between $\left\{ 0,1\right\} $. If $x<P$
then $E^{n}$ is accepted as the new system state and $E^{s}(\mathbf{f},\mathbf{d})=E^{n}(\mathbf{f^{n}},\mathbf{d^{n}})$.
Notice from Eq. \ref{eq:prob} that if $E^{n}<E^{s}$, then the neighboring
state is always accepted as the new system state. The process of calling
neighbor states and deciding whether or not to accept them as system
state is repeated over $C$ cycles. This represents a random walk
of $C$-steps through the $\left\{ \mathbf{f,d}\right\} $ parameter
space. For optimization calculations performed here, $C$ was chosen
to be $500$ as for values greater than this no appreciable difference
in the results was observed. At each iteration of the algorithm, only
values in the dilute regime and outside the realm of quantum effects
were accepted (i.e. $\sum f_{n}\leq0.1$ and $d_{n}\geq10$ nm were
enforced at each step). In addition, the maximum particle size was
constrained to $d_{n}\leq30$ nm as mentioned above and only physically
realizable values of volume fractions were permitted ($f_{n}\geq0$).
If the algorithm suggested a neighbor state outside the allowable
domain, a large value was assigned to the energy ($E^{n}\sim10^{8}$).
This, by virtue of Eq. \ref{eq:prob}, gives a very low probability
for such states to be accepted. Finally, $T$ is reduced and the entire
process repeated again. $T$ is reduced according to a prescribed
schedule until it is nearly equal to zero after $N$ iterations. Notice
that as $T$ is slowly reduced the system begins to accept lower and
lower energy configurations until the it is forced to into a (global)
minimum. For calculations conducted here, the annealing schedule prescribed
was $T_{N}=(1-\mu)T_{N-1}$ where $\mu=0.99$ using $N=1000$ iterations.
The initial value of $T$ was chosen so that the probability of the
algorithm proceeding from a lower state to a higher state and vice
versa was approximately the same. This ensured that the search space
was relatively large during the initial stages of the anneal. A value
of $T=100$ with $k=0.1$ was used for simulations here. Typically,
$\mathbf{f=}<3.0\%,3.0\%,3.0\%>$ and $\mathbf{d=}<15.0nm,15.0nm,15.0nm>$
was taken as an initial guess. When numerical parameters were determined
such that different simulation runs yielded the same answer (within
a tolerance of $\Delta E\sim10^{-7}$), different initial states where
chosen and simulations ran again. This was done to verify that the
algorithm had, in fact, determined the system configuration yielding
the global minimum. The optimum configuration was determined to be
5.45\% Cu, 0.421\% Al, and 4.11\% Ag in Si$_{3}$N$_{4}$ with respective
nanosphere diameters of 30 nm, 22.27 nm, and 10.0 nm. The spectral
response is show in Fig. \ref{figAppl}b. We note that configurations
of this type do exhibit some enhancement to the complex index of refraction
($N_{eff}$) as compared to the index of the refraction of the stand-alone
Si$_{3}$N$_{4}$ matrix (e.g., Figs. \ref{fig:CxEeffNeff}(a) and
\ref{fig:CxEeffNeff}(b) present the real and imaginary portions of
$\epsilon_{eff}$ and $N_{eff}$ for a Cu:Al:Ag in Si$_{3}$N$_{4}$
nanocomposite system respectively), leading to enhanced reflectivity
at this wavelengths lessening the enhancement of the increased absorption.
However, it is common practice to compensate for this effect by introducing
an anti-reflection layer on top of the solar absorption coating \cite{BostromSolarEngMat2004}.
Similarly, the data for an optimized quaternary nanocomposite of Cu:Ag:Au-SiO$_{2}$
from \cite{TriceSPIE07a} is presented in Fig. \ref{figAppl}(b).
The optimal configuration determined in that study was 1.80\% Cu,
0.35\% Ag, and 6.40\% Au with respective particle diameters of 10.1
nm, 29.6 nm and 10.0 nm. Both absorbing layers may have applications
in enhancing the photocurrent of existing photovoltaic devices.

\section{Conclusions and Discussion \label{sec:Conclusions}}

Here we have applied the mixing rule developed by Garcia and co-workers
\cite{GarciaPRB07} to model the composite electromagnetic behavior
for configurations of Al, Au, Ag, and Cu metallic nanospheres embedded
in dielectric hosts consisting of Si$_{3}$N$_{4}$, TiO$_{2}$, and
SiO$_{2}$. The presence of metal nanoparticles greatly enhances the
absorption spectrum predicted from that of the standalone dielectric
matrix. Certain combinations of metal and nanoparticle concentrations
exhibited a broadband response (e.g. Cu in Si$_{3}$N$_{4}$, Al in
TiO$_{2}$, and Cu in SiO$_{2}$ ) while strong localized absorption
was predicted for other systems (e.g. Ag in all mentioned host dielectrics).
Of note, the resonance absorption enhancement predicted for Au, Ag,
and Cu in Si$_{3}$N$_{4}$ and TiO$_{2}$ all correspond to photon
energies below the respective bandgaps of the host dielectric ( 4.5
eV for Si$_{3}$N$_{4}$ and 3.0 eV for TiO$_{2}$). Such interaction
may have application in carrier generation for the next generation
of photocatalysis and semiconductor devices. In addition, potential
uses for spectral sensing and an optimization approach to find preferential
system combinations for solar energy harvesting have been presented
and discussed. The use of such nanocomposite systems as absorbing
layers on photovoltaic devices adds another degree of freedom for
device tailoring and potential efficiency enhancement. Consideration
of the stability of such embedded nanoparticle systems under various
processing conditions and the steps towards fabrication of nanocomposites
of the type described are currently under investigation.

RK, and RS acknowledge support by the National Science Foundation
through grants \# CMMI-0855949, while HG acknowledges support by the
National Science Foundation through grant \# CMMI-0757547.

\pagebreak\bibliographystyle{elsart-num}

\pagebreak

\pagebreak

\pagebreak

\end{document}